\documentclass[aps,prb,twocolumn]{revtex4}
\usepackage{graphicx,amssymb}
\usepackage[usenames,dvipsnames]{pstricks}
\usepackage{epsfig} 
\usepackage{pst-grad} 
\usepackage{pst-plot} 

\begin{document}

\title{J.S. Bell's Concept of Local Causality}

\author{Travis Norsen}
\affiliation{Marlboro College, Marlboro, VT  05344} 
\email{norsen@marlboro.edu}

\date{March 3, 2011}

\begin{abstract}

John Stewart Bell's famous 1964 theorem is widely regarded as one of the most
important developments in the foundations of physics.  It has even been
described as ``the most profound discovery of science.''  Yet even as
we approach the 50th anniversary of Bell's discovery, its meaning and
implications remain controversial.  Many textbooks and commentators
report that Bell's theorem refutes the possibility (suggested
especially by Einstein, Podolsky, and Rosen in 1935) of supplementing 
ordinary quantum theory with additional (``hidden'') variables that 
might restore determinism and/or some notion of an
observer-independent reality.  On
this view, Bell's theorem supports the orthodox Copenhagen
interpretation.  Bell's own view of his theorem, however, was 
quite different.  He
instead took the theorem as establishing an ``essential conflict''
between the now well-tested empirical predictions of quantum
theory and relativistic \emph{local causality}.
The goal of the present paper is, in general, to make Bell's own views 
more widely known and, in particular, to explain in detail Bell's
little-known mathematical formulation of the concept of relativistic local
causality on which his theorem rests.  
We thus collect and organize many of Bell's
crucial statements on these topics, which are scattered throughout his 
writings, into a self-contained, pedagogical discussion including
elaborations of the concepts ``beable'', ``completeness'', and ``causality''
which figure in the formulation.  We also show how local
causality (as formulated by Bell) can be used to derive an empirically
testable Bell-type inequality, and how it can be used to recapitulate
the EPR argument.
\end{abstract}

\maketitle

\section{Introduction}

In its most generalized sense, ``local causality'' is the idea that 
physical influences propagate continuously through space -- i.e., the
idea that what Einstein famously called ``spooky actions at a
distance'' are impossible.\cite{bel}  In addition to originating
this catchy
slogan, Einstein was also chiefly responsible for the modern, specifically
relativistic sense of local causality, according to which causal
influences should not only propagate continuously (never hopping
across a gap in which no trace is left) but should do so always at the
speed of light or slower.  The elaboration and precise formulation of
this idea will be our central concern here. 

It is interesting to first note, though, that the pre-relativistic 
``no action at a distance'' sense of local causality has played an
important  role in the construction and assessment of theories
throughout 
the history of physics.\cite{hesse,mcmullin}  For example, some
important objections to Isaac Newton's theory of gravitation 
centered on the theory's alleged positing of such
non-local action at a distance.  Newton's own view, interestingly, 
seems to have been
that although his theory did claim that (for example) the sun exerted
causal influences on the distant planets, this was in principle
perfectly consistent with local causality, which he strongly endorsed:
\begin{quote}
``It is inconceivable that inanimate brute matter should, without the
mediation of something else which is not material, operate upon and
affect other matter without mutual contact... That gravity should be
innate, inherent, and essential to matter, so that one body may act
upon another at a distance through a vacuum, without the mediation of
anything else, by and through which their action and force may be
conveyed from one to another, is to me so great an absurdity that I
believe no man who has in philosophical matters a competent faculty of
thinking can ever fall into it.''\cite{bentleyletter}
\end{quote}
Newton's idea\cite{cohen}
was evidently that his gravitational theory simply didn't
yet provide a complete description of the underlying (and presumably
local) mechanism ``by and through which [massive bodies'] action and 
force may be conveyed from one to another.'' 

Such debates had a philosophical character, though, 
since at the time nothing 
was definitely, unambiguously excluded by the
requirement of locality.  \emph{Any} apparent action at a distance in a 
theory could be rendered compatible in principle with local causality 
by following Newton and simply denying that the theory in question 
provided a \emph{complete} description of the relevant phenomena.  

This changed in 1905 with Albert Einstein's discovery of Special
Relativity (SR), which for the first time purported to identify 
a certain class of
causal influences -- namely, those which propagate faster than light --
as definitely inconsistent with local causality.
As Einstein explained,
\begin{quote}
``The success of the
Faraday-Maxwell interpretation of electromagnetic action at a distance
resulted in physicists becoming convinced that there are no such
things as instantaneous action at a distance (not involving an
intermediary medium) of the type of Newton's law of gravitation.
According to the theory of relativity, action at a distance with the
velocity of light always takes the place of instantaneous action at a
distance or of action at a distance with an infinite velocity of
transmission.  This is connected with the fact that the velocity $c$
plays a fundamental role in this theory.''\cite{einstein}
\end{quote}
The speed of light $c$ plays a fundamental role vis-\`a-vis 
causality in SR because of the relativity of simultaneity.  For two
events $A$ and $B$ with space-like separation (i.e., such that a
signal connecting $A$ and $B$ would have to propagate
faster than $c$), the time ordering is ambiguous:  different inertial
observers will disagree about whether $A$ precedes $B$ in time, or
vice versa.  There is thus, according to SR, no objective matter of
fact about which event occurs first -- and hence no possibility of a
causal relation between them, since the relation between a cause and
its effect is necessarily time-asymmetric. 
As J.S. Bell put this point, ``To avoid causal
chains going backward in time in some frames of reference, we require
them to go slower than light in any frame of reference.''\cite{bell1990} 

After the advent of SR, it didn't take long for the relativistic
sense of local causality to be deployed in a criticism of
other developing theories, much as the pre-relativistic concept had
been used against Newton's theory.  Indeed, it was Einstein himself -- in both
the famous, but widely misunderstood, EPR paper\cite{epr} and several
related but less widely known arguments\cite{boxes} -- who first
pointed out that the developing Copenhagen quantum theory violated SR's
locality constraint.  In
particular, according to Einstein, that theory's account of 
\emph{measurement} combined with Niels Bohr's \emph{completeness
doctrine} committed the theory to precisely the sort of non-local
causation that was, at least according to Einstein, prohibited by SR.  
Einstein thus rejected Bohr's completeness
doctrine and supported (something like) what is now 
(unfortunately\cite{fn:hvs}) called the local ``hidden variables'' 
program.  

Note the interesting parallel to Newtonian gravity here, with the 
non-locality in some candidate theory 
being rendered either real or merely apparent,
depending on whether or not one interprets the theory as providing 
a \emph{complete} description of the physical processes in question.
Einstein's assessment of Copenhagen  quantum theory with respect to  
local causality is thus logically parallel to Newton's
analysis of his own theory of gravitation:  the theory, if regarded as
complete, violates locality -- hence upholding locality requires denying
completeness.  

This brings us to the main
subject of the present paper: the work of J.S. Bell.  Bell (unlike 
many commentators) accepted 
Einstein's proof of the non-locality of Copenhagen quantum theory.  In
particular, Bell accepted as valid ``the EPR argument 
\emph{from locality to} deterministic hidden variables.''\cite{bell1981b}
The setup for this argument involves a pair of specially-prepared
particles which are allowed to separate to remote 
locations.  An observation of some property of one particle then
permits the
observer to learn something about a corresponding property of the
distant particle.  According to the Copenhagen view, the distant
particle fails to possess a definite value for 
the property in question prior to the
observation, and so it is precisely the observation of the nearby
particle which  -- in apparent violation of local causality -- 
triggers the crystallization of this newly real property for the
distant particle.

In Bell's recapitulation of the argument, though, 
for Einstein, Podolsky, and Rosen (EPR) this
\begin{quote}
``simply showed that [Bohr, Heisenberg, and Jordan] had been
hasty in dismissing the reality of the microscopic world.  In
particular, Jordan had been wrong in supposing that nothing was real
or fixed in that world before observation.  For after observing only
one particle the result of subsequently observing the other (possibly
at a very remote place) is immediately predictable.  Could it be that
the first observation somehow fixes what was unfixed, or makes real
what was unreal, not only for the near particle but also for the
remote one?  For EPR that would be an unthinkable `spooky action at a
distance'.  To avoid such action at a distance [one has] to attribute,
to the space-time regions in question, \emph{real} properties in
advance of observation, correlated properties, which
\emph{predetermine} the outcomes of these particular observations.
Since these real properties, fixed in advance of observation, are not
contained  in quantum formalism, that formalism for EPR is
\emph{incomplete}.  It may be correct, as far as it goes, but the
usual quantum formalism cannot be the whole story.''  \cite{bell1981b}
\end{quote}
Bell thus agreed with 
Einstein that the local hidden variables program constituted
the \emph{only hope} for a locally causal re-formulation of quantum theory.  

Bell's historic contribution, however,
was a 1964 theorem establishing that no such local hidden variable theory 
-- \emph{and hence no local theory of any kind} -- could
generate the correct empirical predictions for a certain class of
experiment.\cite{bell1964}
According to Bell, we must therefore
accept the real existence, in nature, of faster-than-light causation
-- in apparent conflict with the requirements of SR:
\begin{quote}
``For me then this is the real problem with quantum theory:  the
apparently essential conflict between any sharp formulation and
fundamental relativity.  That is to say, we have an apparent
incompatibility, at the deepest level, between the two fundamental
pillars of contemporary theory...''  \cite{bell1984}
\end{quote}
Indeed, Bell even went so far as to suggest, in response to his theorem
and the relevant experimental data,\cite{aspect,weihs} 
the rejection of ``fundamental
relativity'' and the return to a Lorentzian view in which there is a
dynamically privileged (though probably empirically undetectable)
reference frame:
\begin{quote}
``It may well be that a relativistic version of [quantum]
theory, while Lorentz invariant and local at the observational level,
may be necessarily non-local and with a preferred frame (or aether) at
the fundamental level.''\cite{bell1981}
\end{quote}
And elsewhere: 
\begin{quote}
``...I would say that the cheapest resolution is something like going
back to relativity as it was before Einstein, when people like Lorentz
and Poincar\'e thought that there was an aether -- a preferred frame of
reference -- but that our measuring instruments were distorted by
motion in such a way that we could not detect motion through the
aether.  Now, in that way you can imagine that there is a preferred
frame of reference, and in this preferred frame of reference things do
go faster than light.  .... Behind the apparent Lorentz invariance of
the phenomena, there is a deeper level which is not Lorentz
invariant...  [This] pre-Einstein position of Lorentz and Poincar\'e,
Larmor and Fitzgerald, was perfectly coherent, and is not inconsistent
with relativity theory.  The idea that there is an aether, and these
Fitzgerald contractions and Larmor dilations occur, and that as a
result the instruments do not detect motion through the aether -- that
is a perfectly coherent point of view.''\cite{ghostatom,fn:sr}  
\end{quote}
Here our intention is not to lobby for this radical view, but simply to
explain Bell's rationale for contemplating it.  

This rationale
involves a complex chain of reasoning including at least these four steps:
(i) arguing that SR prohibits causal influences between space-like
separated events, (ii) constructing a mathematically precise
formulation of this prohibition, i.e., of relativistic local
causality, (iii) deriving an empirically-testable inequality from this
formulation of local causality, and then (iv) establishing that the
inequality is inconsistent with actual empirical data.  There is an
extensive ``Bell literature'' in which each of these steps is
subjected to probing critical analysis.  The time-asymmetric 
character of causal relations -- used in the argument
for (i) that was sketched above -- has for example been challenged by
Huw Price\cite{price} and, in a rather different way, 
by recent work of Roderich Tumulka\cite{tumulka} (which was, incidentally,
based on earlier work by Bell\cite{bell1987}).   
And there remain certain ``loopholes'' in the experiments 
demonstrating violations of Bell's inequality, such that one might 
conceivably doubt (iv).\cite{experiments}  

For the most part, though, physicists do not seriously question (i)
and regard (iv) as established with reasonable conclusiveness.  The
controversies about the meaning and implications of Bell's theorem
have thus centered on points (ii) and (iii).  Clearly, though, what
one says about point
(iii) -- the question of whether and how a Bell-type inequality is
entailed by local causality --  will depend strongly on whether and
how one has addressed (ii).  And sadly, Bell's own views in regard to
(ii) have been almost entirely invisible in the Bell literature.  (The
review article cited in Ref.~\onlinecite{experiments}, for example,
doesn't acknowledge Bell's formulation of local causality at all but 
instead proposes an alternative formulation very different from
Bell's.)  It is thus not terribly surprising that so many commentators
and textbook authors have summarized the upshot of Bell's theorem in
ways so different from Bell's own.  Typically, for example, one encounters
the claim that Bell's inequality follows not from local causality
alone, but from the conjunction of local causality with some
additional premises;  some of the usual suspects here include 
``hidden variables,'' ``determinism,''
``realism,'' ``counter-factual definiteness'', or an improper
insistence on a vaguely-defined ``classical'' way of thinking.  
One or more of these (rather than relativistic
local causality) is then invariably blamed for the inconsistency with
experiment.\cite{wigner,mermin,jarrett-cm,zeilinger,gz,griffiths,townsend,
sakurai}

A full presentation of Bell's alternative to these widespread views
would include a systematic treatment of point (iii).  We will sketch
this derivation in Section \ref{sec6}, but the bulk of our discussion
will focus instead on Bell's formulation of local causality, i.e., his
views on point (ii).  This discussion will be based
primarily on (but will also in several ways interpret and extend
beyond) Bell's 1990 paper ``La nouvelle cuisine'' (published in the
same year as his untimely death).  Clearly explaining Bell's
formulation of locality, however, will require also sketching Bell's
interesting and refreshingly unorthodox views on a number of related
issues in the foundations of quantum theory.  The discussion will 
thus be elaborated and supported with excerpts from Bell's many 
other papers. 

The main audience for the paper is students and physicists with little
or no prior knowledge of Bell's theorem beyond what they've read in
textbooks.  It should be understood, though, that virtually all of 
the issues raised 
here are included because some kind of misunderstanding
(or simple ignorance) of
them has been present and influential in the Bell literature.  We will
provide occasional citations to works which we think exemplify the various
important misunderstandings.  But length considerations (and the
desire to keep this a self-contained, positive presentation of Bell's
views) forbid any extensive polemical discussions.  
Still, those familiar with the Bell literature will have no
trouble appreciating where Bell's own views challenge the conventional
wisdom.  It is hoped that simply bringing Bell's views more out into the
open will stimulate fruitful debates among those with interests in
these areas.

The paper is organized as follows.  In the following section, we jump
quickly from some of Bell's preliminary, qualitative statements 
to his final, quantitative formulation of relativistic local
causality.  Then, in Sections \ref{sec3}-\ref{sec5}, 
we will highlight and explore various aspects by clarifying some
perhaps-unfamiliar or suspicious terms which appear in Bell's 
formulation and by contrasting them to various other ideas with
which they have sometimes been confused.  Section \ref{sec6} will
show how local
causality (as formulated by Bell) can be used to derive an empirically
testable Bell-type inequality, and also how it can be used to recapitulate
the EPR argument.  Finally, in Section \ref{sec7} we will summarize
the arguments presented and acknowledge some limitations of and open
questions about Bell's formulation.

\begin{figure}[t]
\begin{center}
\scalebox{.7}{
\scalebox{1} 
{
\begin{pspicture}(0,-2.64)(10.04,2.64)
\pscircle[linewidth=0.04,dimen=outer](3.3,0.0){0.5}
\usefont{T1}{ptm}{m}{n}
\rput(3.2914062,0.01){$1$}
\psline[linewidth=0.04cm](3.58,0.4)(6.58,-2.6)
\psline[linewidth=0.04cm](3.0,0.38)(0.0,-2.62)
\pscircle[linewidth=0.04,dimen=outer](8.4,-0.02){0.5}
\usefont{T1}{ptm}{m}{n}
\rput(8.391406,-0.01){$2$}
\psline[linewidth=0.04cm](0.0,2.62)(3.0,-0.38)
\psline[linewidth=0.04cm](6.6,2.62)(3.6,-0.38)
\usefont{T1}{ptm}{m}{n}
\rput(3.2914062,1.21){$\text{effects}$}
\usefont{T1}{ptm}{m}{n}
\rput(3.2814062,-1.27){$\text{causes}$}
\usefont{T1}{ptm}{m}{n}
\rput(8.401406,1.11){$\text{time}$}
\usefont{T1}{ptm}{m}{n}
\rput(8.4114065,-1.25){$\text{space}$}
\psline[linewidth=0.04cm,arrowsize=0.05291667cm 2.0,arrowlength=1.4,arrowinset=0.4]{<-}(6.92,-1.26)(7.72,-1.26)
\psline[linewidth=0.04cm,arrowsize=0.05291667cm 2.0,arrowlength=1.4,arrowinset=0.4]{<-}(10.02,-1.26)(9.22,-1.26)
\end{pspicture} 
}
}
\caption{
``Space-time location of causes and effects of events in region 1.''
(Figure and caption are from Ref.~\onlinecite{bell1990}.)
\label{fig1}
}
\end{center}
\end{figure}
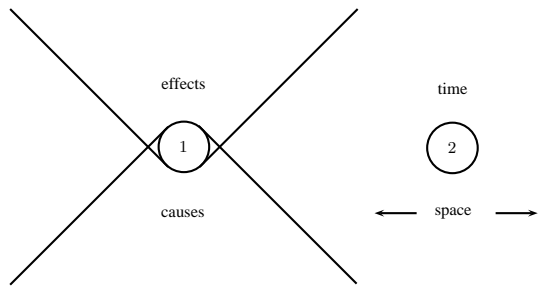

\section{Local Causality: Overview}
\label{sec2}

Let us begin with a qualitative formulation of Bell's concept of local
causality.  In a 1988 interview, in answer to the
question ``What does \emph{locality} mean?'' Bell responded:\cite{fn:locvslc}
\begin{quote}
``It's the idea that what you do has consequences only nearby, and
that any consequences at a distant place will be weaker and will
arrive there only after the time permitted by the velocity of light.
Locality is the idea that consequences propagate continuously, that
they don't leap over distances.''\cite{omni}  
\end{quote}
Bell gave a slightly more careful (but still qualitative) formulation
of what he called the ``Principle of local causality'' in his 1990 paper:
\begin{quote}
``The direct causes (and effects) of events are near by, and even the
indirect causes (and effects) are no further away than permitted by
the velocity of light.''\cite{bell1990}
\end{quote}
Then, citing a figure which we have reproduced here as Figure
\ref{fig1}, Bell continues:  
\begin{quote}
``Thus for events in a space-time region 1 ... we would look for causes
in the backward light cone, and for effects in the future light cone.
In a region like 2, space-like separated from 1, we would seek neither
causes nor effects of events in 1.''\cite{bell1990}
\end{quote}
This should be relatively uncontroversial.
Bell immediately notes, however, that ``[t]he above principle of local
causality is not yet sufficiently sharp and clean for
mathematics.''\cite{bell1990}

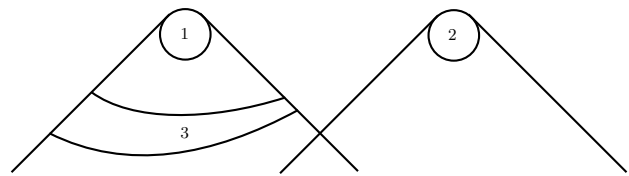
\begin{figure}[t]
\begin{center}
\scalebox{.7}{
\scalebox{1} 
{
\begin{pspicture}(0,-1.59)(11.7,1.57)
\pscircle[linewidth=0.04,dimen=outer](3.3,1.07){0.5}
\usefont{T1}{ptm}{m}{n}
\rput(3.2914062,1.08){$1$}
\psline[linewidth=0.04cm](3.58,1.47)(6.58,-1.53)
\psline[linewidth=0.04cm](3.0,1.45)(0.0,-1.55)
\pscircle[linewidth=0.04,dimen=outer](8.4,1.05){0.5}
\usefont{T1}{ptm}{m}{n}
\rput(8.371407,1.06){$2$}
\psline[linewidth=0.04cm](8.68,1.45)(11.68,-1.55)
\psline[linewidth=0.04cm](8.1,1.43)(5.1,-1.57)
\psbezier[linewidth=0.04](1.52,-0.03)(2.28,-0.59)(3.76,-0.59)(5.2,-0.13)
\psbezier[linewidth=0.04](0.72,-0.81)(2.2,-1.55)(3.82,-1.25)(5.44,-0.37)
\usefont{T1}{ptm}{m}{n}
\rput(3.2914062,-0.8){$3$}
\end{pspicture} 
}
}
\caption{
``Full specification of what happens in 3 makes events in 2 irrelevant
for predictions about 1 in a locally causal theory.''  (Figure and
caption are from Ref.~\onlinecite{bell1990}).
\label{fig2}
}
\end{center}
\end{figure}

Here, then, is Bell's sharpened and cleaned formulation of
special relativistic local causality.  (The reader should remember
that this is, at this point, merely a `teaser' which those to whom it
is not already familiar should only expect to understand after
further reading.)
\begin{quote}
``A theory will be said to be locally causal 
if the probabilities attached to values of local beables
in a space-time region 1 are unaltered by specification of values of
local beables in a space-like separated region 2, when what happens in
the backward light cone of 1 is already sufficiently specified, for
example by a full specification of local beables in a space-time
region 3...''  \cite{bell1990}
\end{quote}
The space-time regions referred to are illustrated in Figure 2.
We may translate Bell's formulation into mathematical form as follows:
\begin{equation}
P(b_1 | B_3, b_2) = P(b_1 | B_3),
\label{locality}
\end{equation}
where $b_i$ refers to 
the value of some particular beable in space-time region $i$
and $B_i$ refers to a \emph{sufficient} (for example, a complete) 
specification of \emph{all}  beables in the
relevant region. 
The $P$s here are the probabilities assigned to event $b_1$ by the
candidate theory in question.
Eq.~(\ref{locality}) thus 
asserts mathematically just what Bell states in the caption of his
accompanying figure (reproduced here as Figure \ref{fig2}):  
``full specification of [beables] in 3 makes events in 2 
irrelevant for predictions about 1 in a locally causal
theory.''\cite{bell1990}

Let us then jump right in to a closer examination
of the several perhaps-puzzling features of this formulation.

\section{Beables}
\label{sec3}

For those to whom the term is new, the first question about the
word ``beable'' is:  how to pronounce it?  The word has three
syllables.  It does not rhyme with ``feeble,'' but with ``agreeable.''   
Bell invented the
word as a contrast to the ``observables'' which play a fundamental
role in the formulation of orthodox quantum theory, so let us begin 
there.

\subsection{Beables vs. Observables}

Beables (as
contrasted to observables) are those elements of a theory which
are supposed to
correspond to something that is \emph{physically real}, independent of
any observation.  Bell elaborates:
\begin{quote}
``The beables of the theory are those elements which might
correspond to elements of reality, to things which exist.  Their 
existence does not depend on `observation'.  Indeed observation
and observers must be made out of beables.'' \cite{bell1984b}
\end{quote}
Or as he explains elsewhere,
\begin{quote}
``The concept of `observable' .... is a rather woolly concept.  It is
not easy to identify precisely which physical processes are to be
given the status of `observations' and which are to be relegated to
the limbo between one observation and another.  So
it could be hoped that some increase in precision might be
possible by concentration on the \emph{be}ables
... because they are there.'' \cite{bell1976}
\end{quote}
Bell's reservations here (about the concept ``observable'' appearing 
in the fundamental formulation of allegedly fundamental theories)
are closely
related to the so-called ``measurement problem'' of orthodox quantum
mechanics, which Bell encapsulated by remarking that the
orthodox theory is ``unprofessionally vague and
ambiguous''\cite{bell1984b} in so far as its fundamental dynamics is 
expressed in terms of
``words which, however legitimate and necessary in application, have
no place in a \emph{formulation} with any pretension to physical
precision'' -- such words as ``\emph{system, apparatus, environment, 
microscopic, macroscopic, reversible, irreversible, observable, 
information, measurement.}''  \cite{bell1989}  As Bell elaborates,
\begin{quote}
``The concepts `system', `apparatus', `environment', immediately imply
an artificial division of the world, and an intention to neglect, or
take only schematic account of, the interaction across the split.  The
notions of `microscopic' and `macroscopic' defy precise definition.
So also do the notions of `reversible' and `irreversible'.  Einstein
said that it is theory which decides what is `observable'.  I think he
was right -- `observable' is a complicated and theory-laden business.
Then the notion should not appear in the \emph{formulation} of
fundamental theory.'' \cite{bell1989}
\end{quote}
As Bell points out, even Bohr (a convenient personification of
skepticism regarding the physical reality of unobservable
microscopic phenomena) recognizes certain things (for example, the
directly perceivable states of a classical measuring apparatus) as
unambiguously real, i.e., as beables:
\begin{quote}  
``The terminology, \emph{be}-able as against \emph{observ}-able,
is not designed to frighten with metaphysic those dedicated to
realphysic.  It is chosen rather to help in making explicit some
notions already implicit in, and basic to, ordinary quantum theory.
For, in the words of Bohr, `it is decisive to recognize that, however
far the phenomena transcend the scope of classical physical
explanation, the account of all evidence must be expressed in
classical terms.'  It is the ambition of the
theory of local beables to bring these `classical terms' into the
equations, and not relegate them entirely to the surrounding talk.''
\cite{bell1976}
\end{quote}  
The unprofessional vagueness and ambiguity of
orthodox quantum theory, then, is related to the fact that its
formulation presupposes these (classical, macroscopic) beables, 
but fails to provide clear mathematical laws to describe them.  
As Bell explains, 
\begin{quote}
``The kinematics of the
world, in [the] orthodox picture, is given by a wavefunction ... for
the quantum part, and classical variables -- variables which
\emph{have} values -- for the classical part... [with the classical
variables being] somehow macroscopic.  This is not spelled out very
explicitly.  The dynamics is not very precisely formulated either.  It
includes a Schr\"odinger equation for the quantum part,  and some sort
of classical mechanics for the classical part, and `collapse' recipes
for their interaction.''  \cite{bell1989}
\end{quote}
There are thus two related problems.  First,
the posited ontology is rather different on the two sides of (what
Bell calls) ``the shifty split'' \cite{bell1989}
 -- that is, the division between
``the quantum part'' and ``the classical part.''  
But then, as a whole, the posited ontology remains
unavoidably vague so long as the split remains shifty -- i.e., so long
as the dividing line between the macroscopic and microscopic remains
undefined.  And second,
the \emph{interaction} across the split is
problematic.  Not only is the account of this dynamics (the
``collapse'' process) inherently bound up in concepts from Bell's list
of dubious terms, but the very existence of a special dynamics for the
interaction seems to imply inconsistencies with the
dynamics already posited for the two realms separately.  

As Bell summarizes, 
\begin{quote}
``I think there are \emph{professional} problems
[with quantum mechanics].  That is to say, I'm a professional
theoretical physicist and I would like to make a clean theory.  And
when I look at quantum mechanics I see that it's a dirty theory.
The formulations of quantum mechanics that you find in the books
involve dividing the world into an observer and an observed, and
you are not told where that division comes... So you have a theory
which is fundamentally ambiguous...'' \cite{ghostatom}
\end{quote}
The point of all this is to clarify the sort of theory Bell
had in mind as satisfying the relevant standards of professionalism in
physics.  It is often thought, by those who do not understand or do
not accept Bell's criticisms of orthodox quantum theory, that the very
concept of ``beable'' (in terms of which his concept of local
causality is formulated) commits one already to hidden variables or
determinism or some sort of naive realism or some other 
physically or philosophically dubious principle.  

But this is not correct.  The requirement here,
ultimately, is only that candidate fundamental theories -- at least,
those ``with any pretension to physical
precision'' \cite{bell1989} -- be formulated \emph{clearly} and
\emph{precisely}.  And this requires, according to Bell, that the theories
provide a \emph{uniform} and \emph{consistent} candidate description 
of physical reality.  In particular, there should be no ambiguity
or inconsistency regarding what a given candidate theory is fundamentally
\emph{about} (the beables), nor regarding precisely how those 
posited physically real elements are posited to act and interact 
(the laws).

\subsection{Beables vs. Conventions}

So far we have explained the term ``beable'' by contrasting it to the
``observables'' of orthodox quantum theory. 
We must now also contrast the concept of ``beables'' with
those elements of a theory which are, to some degree, conventional:  
\begin{quote}
``The word `beable' will also be used here to carry another
distinction, that familiar already in classical theory between
`physical' and `non-physical' quantities.  In Maxwell's
electromagnetic theory, for example, the fields 
${\bf{E}}$ and ${\bf{H}}$ are `physical'
(beables, we will say) but the potentials ${\bf{A}}$ and $\phi$ are
`non-physical'.  Because of gauge invariance the same physical
situation can be described by very different potentials.  It does not
matter [i.e., it is not a violation of local causality] that in 
Coulomb gauge the scalar potential propagates with
infinite velocity.  It is not really supposed to \emph{be} there.  It
is just a mathematical convenience.''\cite{bell1976}
\end{quote}
Or, as Bell puts the same point in another paper,
\begin{quote}
``...there \emph{are} things which \emph{do} go faster than light.
British sovereignty is the classical example.  When the Queen dies in
London (long may it be delayed) the Prince of Wales, lecturing on
modern architecture in Australia, becomes \emph{instantaneously}
King....  And there are things like that in physics.  In Maxwell's
theory, the electric and magnetic fields in free space satisfy the
wave equation 
\begin{eqnarray}
\frac{1}{c^2} \frac{\partial^2 {\bf{E}}}{\partial t^2} -
{\bf{\nabla}}^2 {\bf{E}} &=& 0 \nonumber \\
\frac{1}{c^2} \frac{\partial^2 {\bf{B}}}{\partial t^2} -
{\bf{\nabla}}^2 {\bf{B}} &=& 0 \nonumber
\end{eqnarray}
...corresponding to propagation with velocity $c$.  But
the scalar potential, if one chooses to work in `Coulomb gauge',
satisfies Laplace's equation 
\begin{eqnarray}
- {\bf{\nabla}}^2 \phi = 0 
\nonumber
\end{eqnarray}
...corresponding to propagation with
infinite velocity.  Because the potentials are only mathematical
conveniences, and arbitrary to a high degree, made definite only by
the imposition of one convention or another, this infinitely fast
propagation of the Coulomb-gauge scalar potential disturbs no one.
Conventions can propagate as fast as may be convenient.  But then we
must distinguish in our theory between what is convention and what is
not.''  \cite{bell1990}
\end{quote}
Thus, in order to cleanly decide whether a given theory is or is not
consistent with local causality,
\begin{quote}
``you must identify in your theory `local \emph{be}ables'.  The
\emph{be}ables of the theory are those entities in it which are, at
least tentatively, to be taken seriously, as corresponding to
something real.  The concept of `reality' is now an embarrassing one
for many physicists....  But if you are unable to give some special
status to things like electric and magnetic fields (in classical
electromagnetism), as compared with the vector and scalar potentials,
and British sovereignty, then we cannot begin a serious discussion.''
\cite{bell1990}
\end{quote}
This explains why, according to Bell:
``It is in terms of local beables that we can hope to formulate
some notion of local causality.'' \cite{bell1976}

\subsection{Beables and Candidate Theories}

It is important to appreciate that a beable is only a beable
relative to some particular candidate theory which posits those
elements as physically real (and, presumably, gives precise
mathematical laws for their dynamics).  For example, the fields
${\bf{E}}$ and ${\bf{B}}$ (and not the potentials) are beables
according to classical Maxwellian  electrodynamics as it is normally
understood.  But one could
imagine some alternative theory which (perhaps motivated by 
the Aharanov-Bohm effect) posits the Coulomb gauge potentials as beables
instead.  Note that,
although it would be empirically and in some sense mathematically
equivalent to the usual theory, 
the alternative theory would evidently violate local 
causality (because, say, wiggling a charge would instantaneously 
affect the physically-real scalar potential at distant locations)
while the usual, Maxwellian theory would respect it.

Thinking in terms of such candidate theories helps us
separate any questions about what the ``real beables'' are -- what
really exists out there in physical reality -- into
two distinct parts:  first, what elements does a given candidate
theory \emph{posit} as beables; and second, which candidate theory do we
think is \emph{true}?  The point is, you don't have to be able to
answer the second question in order to answer (for a given theory) the
first.  This should provide some comfort to those who (perhaps
influenced by positivist or instrumentalist philosophy) think we can't
(and/or shouldn't try to) establish some theoretical picture of
external reality as true.
Such people may still accept Bell's characterization of when 
``a theory will be said to be locally causal.''

But even those who are not skeptical on principle recognize
that, because of the complexity in practice of settling questions about
the truth status of scientific theories, some tentativeness is 
often in order.  Bell recognizes this too: 
\begin{quote}
``I use the term
`beable' rather than some more committed term like `being' or `beer'
to recall the essentially tentative nature of any physical theory.
Such a theory is at best a \emph{candidate} for the description of
nature.  Terms like `being', `beer', `existent', etc., would seem to
me lacking in humility.  In fact `beable' is short for `maybe-able'.''
\cite{bell1984b}
\end{quote}
The crucial point is that the ``maybe'' here pertains to the
epistemological status of a given candidate theory.  By contrast,
the ``beable
status'' \cite{bell1976}
 of certain elements of a theory \emph{relative to that
theory} should be completely
straightforward and uncontroversial.  If there is any question about
what elements a theory posits as beables, it can only be because the
proponents of the theory have not (yet) sufficiently clarified what
the theory is about, what the theory \emph{is}.
Whether the theory is true
or false is an orthogonal question.

All of that said, Bell does take certain elements largely for granted
as beables -- that is, as beables that \emph{any} serious candidate
theory would have to recognize as such:
``The beables must
include the settings of switches and knobs on experimental equipment,
the currents in coils, and the readings 
of instruments.''\cite{bell1976}  As noted before, even Bohr
acknowledges the real existence (the beable status) of these
sorts of things.  And, as suggested by Bohr, since our primary
cognitive access to the world is through ``switches and knobs on 
experimental equipment'' and other such directly perceivable facts
-- i.e., since such facts must always constitute the primary
\emph{evidence} on which we will have to rest any argument for the
truth of a particular candidate physical theory -- it is hard to
imagine a serious theory which doesn't grant such facts beable
status:  a theory which didn't would evidently have to regard our
perception as systematically delusional, and hence would have to
regard any alleged empirical evidence -- for anything, including
itself -- as invalid.  In short, such a theory would evidently
be self-refuting.  \cite{realism}

We stress this point for two related reasons.  First, anyone who is
uncomfortable with the apparently ``metaphysical'' positing of
ultimate ``elements of reality'' (even in a tentative way, through the
tentative positing of a candidate physical theory) should be relieved
to find that the concept ``beable'' is merely a placeholder for
whatever entities we (tentatively) include in the class which 
\emph{already},
by necessity, exists and 
includes certain basic, directly-perceivable features of
the world around us.  And second, these particular beables -- e.g., the
settings of knobs and the positions of pointers -- have a
particularly central role to play in the derivation (from Bell's
concept of local causality) of the empirically testable Bell
inequalities.  This will be developed in Section \ref{sec6}.

\section{Completeness}
\label{sec4}

Having clarified the concept of ``beables'' which appears in Bell's
formulation of local causality, let us now turn to the last phrase in
that formulation, italicized here:
\begin{quote}
``A theory will be said to be locally causal 
if the probabilities attached to values of local beables
in a space-time region 1 are unaltered by specification of values of
local beables in a space-like separated region 2, \emph{when what happens in
the backward light cone of 1 is already sufficiently specified, for
example by a full specification of local beables in a space-time
region 3...}''  \cite{bell1990}
\end{quote}
In a word, the key assumption here is ``that events in 3 be specified
\emph{completely}'' \cite{bell1990} (emphasis added).

Let us first see why this requirement is necessary.  Consider again
Figure \ref{fig2}, and suppose that $\bar{B}_3$ denotes an
\emph{incomplete} specification of beables in region 3.  It can then
be seen that a violation of 
\begin{equation}
P(b_1 | \bar{B}_3, b_2) = P(b_1 | \bar{B}_3)
\label{eq-incomplete}
\end{equation}
does \emph{not} entail the existence of any super-luminal causal
influences.  For suppose some event ``$X$'' in the overlapping 
backwards light cones of regions 1 and 2 causally influences both 
$b_1$ and $b_2$.  It might then be possible to infer, from $b_2$, 
something about $X$, from which one could in turn infer
something about $b_1$.  Suppose, though, that the incomplete
description of events in region 3 -- $\bar{B}_3$ -- omits precisely the
``traces'' of this past common cause $X$.  Then $b_2$ \emph{could} usefully
supplement $\bar{B}_3$ -- i.e., Eq.~(\ref{eq-incomplete}) could be
violated -- even in the presence of purely local causation.  

Thus, as Bell explains, in order for Eq.~(\ref{locality}) 
to function as a valid locality criterion, 
\begin{quote}
``it is important that events in 3 be specified
  completely.  Otherwise the traces in region 2 of causes of events in
  1 could well supplement whatever else was being used for calculating
  probabilities about 1.  The hypothesis is that any such information
  about 2 becomes redundant when 3 is specified completely.''
  \cite{bell1990} 
\end{quote}
And here is the same point from an earlier paper:
\begin{quote}
``Now my intuitive notion of local
causality is that events in 2 should not be `causes' of events in 1,
and vice versa.  But this does not mean that the two sets of events
should be uncorrelated, for they could have common causes in the
overlap of their backward light cones [in a local theory].  It is
perfectly intelligible then that if [$B_3$] in [region 3] does not contain a
complete record of events in that [region], it can be usefully
supplemented by information from region 2.  So in general it is
expected that [$ P(b_1|b_2,\bar{B}_3) \ne P(b_1|\bar{B}_3)$.]  However, in 
the particular
case that [$B_3$] contains already a \emph{complete} specification of
beables in [region 3], supplementary 
information from region 2 could
reasonably be expected to be redundant.'' \cite{bell1976}
\end{quote}
It is important to stress that, like the concept of beables
itself, the idea of a sufficient (full or complete) specification of
beables is relative to a given candidate theory.  What Bell's local
causality condition requires is that -- \emph{in order to assess the
  consistency between a given candidate theory and the relativistic
  causal structure sketched in Figure \ref{fig1}} -- we must include,
in $B_3$, everything \emph{that candidate theory} says is present (or
relevant) in region 3.  It is not, by contrast, necessary that we
achieve omniscience regarding what \emph{actually exists} in some
spacetime region.  

The appearance of the word ``completeness'' tends to remind
commentators of the EPR argument, and hence apparently also tends to
suggest that Bell smuggled into his definition of local causality the
unwarranted assumption that orthodox quantum theory is incomplete.  (See, for
example, Ref. \onlinecite{chsh}.)  As mentioned earlier, Bell did accept the
validity of the EPR argument.  But this means only that, according to
Bell, local causality -- plus some of QM's empirical predictions -- entail the 
incompleteness of orthodox QM.  His view on that point, however, is no
part of his \emph{formulation} of local causality.  
(See Section \ref{sec6} for further discussion of
the relation between local causality and the EPR argument.)

What Bell's formulation does say is only this:  \emph{whatever}
your theory posits as physically real (in region 3), make sure you 
include all of that when calculating the relevant probabilities to
test whether your theory respects or violates Eq.~(\ref{locality}),
i.e., whether your theory is or isn't locally causal in the sense of
Figure \ref{fig1}.  No assumptions are made about the type of theory
to which the locality criterion can be applied.  In particular, the
incompleteness of orthodox quantum theory (i.e., the existence of
``hidden variables'') is not assumed.  The virtue of Bell's
formulation lies precisely in this generality.

Although it is simplest to understand Bell's local causality condition
as requiring a \emph{complete} specification of beables in some
spacetime region, there is an important reason why Bell explicitly
leaves open the possibility that a specification of ``what happens in
the backward light cone of 1'' might be ``sufficiently specified'' by
something \emph{less} than a complete specification of the beables there.
This has to do with the fact, to be discussed more in Section
\ref{sec6}, that in order to
derive an empirically-testable Bell-type inequality from the local
causality condition, one needs a subsidiary assumption, sometimes
called ``experimental freedom'' or ``no conspiracies''.  This is in
essence the assumption that, in the usual EPR-Bell kind of scenario in
which a central source emits pairs of specially-prepared particles in
opposite directions toward two spatially-separated measuring devices,
it is possible for certain settings on the devices (determining which
of several possible measurements are made on a given incoming
particle) to be made ``freely'' or ``randomly'' -- that is,
\emph{independently of the state of the incident particle pair}.  

In the more recent versions of these experiments, the relevant
settings are made using independent (quantum) random number
generators. \cite{weihs}  According to orthodox QM, there is therefore
nothing in the past light cone of an individual measurement event
foretelling which of the possible measurements will be performed.
But there of course exist alternative candidate theories (such as the
de Broglie - Bohm pilot-wave theory, which is deterministic) according
to which those same settings \emph{are} influenced by events in their
pasts.  But then, the relevant pasts of the device settings
necessarily overlap with the pre-measurement states of the particles
being measured.  A complete specification of beables in the relevant
region containing those pre-measurement states will therefore
inevitably include facts relevant to (if not determining) the device
settings.  And so, in deriving the Bell inequality from local
causality, there is a kind of subtle tension between the requirement
``that events in 3 be specified completely'' \cite{bell1990}
and the requirement that device settings can be made independent of 
the states of the particles-to-be-measured.

To resolve the tension, one need merely allow that the beables in 
the relevant region can be divided up into disjoint classes:  those
which are
influenced by the preparation procedure at the source (and which
thus encode the ``state of the particle pair'') and those which are
to be used in the setting of measurement apparatus parameters.  And
note that these two classes are likely to be far from jointly
exhaustive:  in any plausible candidate theory, there will have to
exist many additional beables (corresponding, for example, to
stray electromagnetic
fields, low energy relic neutrinos, etc.) which are in neither of the
mentioned classes.  One thus expects a considerable ``causal
distance'' between the two classes of beables (at least in a
well-designed experiment).  This makes the ``freedom'' or ``no
conspiracies'' assumption -- namely, the absense of correlations
between the two classes of beables -- quite reasonable to accept.

This issue will be addressed in some more detail in Section
\ref{sec6}.   For now, we simply
acknowledge its existence as a way of explaining why Bell's
formulation of local causality mentions ``complete'' descriptions
of events in region 3 as merely an \emph{example} of the kind of
description which is ``sufficient''.  One might summarize the
discussion here as follows:  what is required for the validity of the
local causality condition is a \emph{complete} specification of
beables in region 3 -- but only those beables
which are \emph{relevant} in some appropriate sense to the event $b_1$
in question in region 1.

\section{Causality}
\label{sec5}

Recall the transition from Bell's preliminary, qualitative formulation
of local causality to the final, ``sharp and clean'' version.  And
recall in particular Bell's statement that the preliminary version was
\emph{insufficiently} sharp and clean for mathematics.  What is it,
exactly, that Bell considered inadequate about the qualitative
statement?  It seems likely that it was the presence there of the
terms ``cause'' and ``effect'' which are notoriously difficult to
define mathematically.  Indeed, about his final formulation Bell 
says:  ``Note, by the way, that 
our definition of locally causal theories, although motivated by talk 
of `cause' and `effect', does not in the end explicitly involve these 
rather vague notions.'' \cite{bell1990}

How exactly does Bell's ``definition of locally causal theories'' 
fail to ``explicitly involve'' the ``rather vague notions'' of cause
and effect?  On its face this sounds paradoxical.  But
the resolution is simple:  what Bell's definition actually avoids
is any specific commitment about what physically exists and how it 
acts.  (Indeed, any such commitments would seriously restrict the
generality of the locality criterion, and hence undermine the
scope of Bell's theorem.)
Instead, Bell's definition shifts the burden of 
providing some definite account of causal processes \emph{to 
theories} and itself merely defines a space-time constraint that
must be met if the causal processes \emph{posited by a candidate
theory} are to be deemed locally causal in the sense of SR.  

The important mediating role of candidate theories vis-\`a-vis 
causality will be further stressed and clarified in the following 
subsection.  Subsequent subsections further clarify
the concept of ``causality'' in Bell's ``local causality'' by
contrasting it with several other ideas with which it has often been
confused or conflated.

\subsection{Causality and candidate theories}

As already discussed, according to Bell 
it is the job of physical theories to 
posit certain physically real structures (beables) and laws governing
their interactions and evolution.  
Thus Bell's definition of locally
causal theories is not a specification of locality for a 
particular type of theory,
namely, those that are ``causal'' -- with the implication that
there would exist also theories that are ``non-causal.'' 
A theory, by the very nature of what we mean by that term in this context,
is automatically causal.  ``Causal theory'' is a redundancy.  And
so, as noted earlier, one must understand Bell's ``definition of 
locally causal theories'' as a criterion that theories -- i.e.,
candidate descriptions of causal processes in nature -- 
must satisfy in order to be in accord with special
relativistic locality.  In short, the causality in the ``definition of locally 
causal theories'' is simply whatever a given candidate theory says,
about whatever it says it about.

As Bell explains, the practical \emph{reason} for defining local
causality in terms of the physical processes posited by some
candidate theory (as opposed to the physical processes which actually
exist in nature) has to do with our relatively direct access to the
one as opposed to the other: 
\begin{quote}
``I would insist here on the distinction between analyzing
various physical theories, on the one hand, and philosophising about
the unique real world on the other hand.  In this matter of causality
it is a great inconvenience that the real world is given to us once
only.  We cannot know what would have happened if something had been
different. We cannot repeat an experiment changing just one variable;
the hands of the clock will have moved, and the moons of Jupiter.
Physical theories are more amenable in this respect.  We can
\emph{calculate} the consequences of changing free elements in a
theory, be they only initial conditions, and so
can explore the causal structure of the theory.  I insist that [the
theory of local beables, i.e., the local causality concept] is
primarily an analysis of certain kinds of physical
theory.''\cite{bell1977}
\end{quote}
Bell's view, contra several commentators \cite{butterfield, brown}, is
thus that no special philosophical account of causation is needed to
warrant the conclusion that violation of the locality condition
implies genuine non-local causation.  For Bell, it is a rather 
trivial matter to decide, for some (unambiguously formulated) candidate
physical theory, what is and is not a genuinely \emph{causal} influence.
We can simply ``explore the causal structure of''
the candidate theory.  This of course raises the question of how we
might go from recognizing the non-locality of some particular
candidate theory, to the claim that \emph{nature} is non-local.  But
this is just Bell's theorem:  \emph{all} candidate theories which
respect the locality condition are inconsistent with experiment.  
(See Section \ref{sec6}.)  So
the ``one true theory'' (whatever that turns out to be!) -- and hence
nature itself -- must violate relativistic local causality.

\subsection{Causality vs. determinism}

The previous subsection stressed that the ``causal'' in ``locally
causal theories'' simply refers to the physically real existents and
processes (beables and associated  laws) posited by some candidate
theory, whatever exactly those might be.  We in no way restrict
the class of theories (whose locality can be assessed by Bell's
criterion) by introducing ``causality.''  In particular, the word 
``causal'' in ``locally causal theories'' is not meant to imply or require
that theories be \emph{deterministic} as opposed to irreducibly
stochastic.  
\begin{quote}
``We would like to form some [notion] 
of \emph{local causality} in  theories which are not deterministic,
in which the correlations prescribed by the theory, for the beables,
are weaker.'' \cite{bell1976}  
\end{quote}
Bell thus uses the word ``causal'' quite deliberately
as a wider abstraction which \emph{subsumes but does not necessarily
entail} determinism.  

This is manifested most clearly 
in the fact that Bell's mathematical formulation of
``local causality'' -- Eq.~(\ref{locality}) -- 
is stated in terms of probabilities.  Indeed, in his 1976 paper 
``The Theory of Local
Beables'' \cite{bell1976} Bell discusses ``Local
Determinism'' \emph{first}, arguing that, in a ``local deterministic''
theory the actual \emph{values} of beables in region 1 (of Figure \ref{fig2})
will be \emph{determined} by a complete specification of beables in region 3
(with additional specification of beables from region 2 being
redundant).  In our mathematical notation, local determinism means
\begin{equation}
b_1(B_3,b_2) = b_1(B_3)
\label{localdeterminism}
\end{equation}
where, as before, $b_1$ and  $b_2$ are the values of specific beables
in regions $1$ and $2$, while $B_3$ denotes a sufficient (e.g.,
complete) specification of beables in region $3$.  

In a (local) 
\emph{stochastic} theory, however, even a complete specification of
relevant beables in the past (e.g., those in region 3 of Figure 2)
may not determine the realized value of the
beable in question (in region $1$).  Rather, the theory specifies only
\emph{probabilities} for the various \emph{possible} values that might
be realized for that beable.  Of course, determinism is not really an
alternative to, but is rather merely a special case of, stochasticity:
\begin{quote}
``Consider for example Maxwell's equations, in the
source-free case for simplicity.  The fields ${\bf{E}}$ and 
${\bf{B}}$ in region 1 are
completely determined by the fields in region 3, regardless of those
in 2.  Thus this is a locally causal theory in the present sense.  The
deterministic case is a limit of the probabilistic case, the
probabilities becoming delta functions.'' \cite{bell1990}
\end{quote}
The natural generalization of the above
mathematical formulation of ``local determinism'' is precisely Bell's
local causality condition:
\begin{equation}
P(b_1 | B_3, b_2) = P(b_1 | B_3),
\label{localcausality2}
\end{equation}
i.e., $b_2$ is irrelevant -- not for determining
\emph{what happens} in region 1 because that, in a stochastic theory,
is simply not determined -- but rather for determining the
\emph{probability} for possible happenings in region 1.  Such probabilities
are the ``output'' of stochastic theories  in the same sense that
the actual realized values of beables are the ``output'' of deterministic 
theories.  Thus, Bell's local causality condition for 
stochastic theories -- Eq.~(\ref{localcausality2}) --
and the analogous condition -- Eq.~(\ref{localdeterminism}) -- for
deterministic theories, are imposing precisely the same locality 
requirement on the two kinds of theories:  information about
region 2 should be irrelevant in 
regard to what the theory says about region 1, once the beables in 
region 3 are sufficiently specified.

Of course, if one insists that any stochastic theory is \emph{ipso
  facto} 
a stand-in for some (perhaps unknown) underlying
deterministic theory (with the probabilities in the stochastic theory
thus resulting not from indeterminism in nature, but from our
ignorance), Bell's locality concept would cease to work.  The requirement
of a complete specification of beables in region 3 would then
contradict the allowance that such a specification does not necessarily
determine the events in region 1.  
But this is no objection
to Bell's concept  of local causality.  Bell is not asking us
to accept that any particular theory (stochastic or otherwise) is
\emph{true}; he's just asking us to accept his definition of what it would
\emph{mean} for a stochastic theory to respect relativity's prohibition on
superluminal causation.  And this requires us to accept, at least in
principle, that there could be such a thing as a genuinely, irreducibly
stochastic theory, and that the way ``causality'' appears in such a
theory is that certain beables do, and others do not, influence the
probabilities for specific events.

We have been stressing here that ``causality'' is wider than, and does
not necessarily entail, determinism.  Bell has deliberately and
carefully formulated 
a local causality criterion that does not tacitly assume determinism,
and which is thus stated explicitly in terms of probabilities  -- 
\emph{the fundamental, dynamical probabilities
assigned by stochastic theories to particular happenings in
space-time.}  Note in particular that 
the probabilities in Eq.~(\ref{locality})
are not subjective (in the sense of denoting the degree of someone's
belief in a proposition about $b_1$), they cannot be understood as 
reflecting partial ignorance about relevant beables in region 3, 
and they do not (primarily) represent
empirical frequencies for the appearance of certain values of $b_1$.  
They are, rather, the fundamental ``output'' of some candidate
(stochastic) physical theory.

\subsection{Causality vs. correlation}

Everyone knows that correlation  doesn't imply causality.  Two events
(say, the values taken on by beables $b_1$ and $b_2$ in Bell's
spacetime regions 1 and 2, respectively) may be correlated without
there necessarily 
being any implication that $b_1$ is the cause of $b_2$ or vice
versa:
\begin{quote}
``Of course, mere \emph{correlation} between distant events does
not by itself imply action at a distance, but only correlation between
the signals reaching the two places.'' \cite{bell1981b}
\end{quote}
And Bell describes the issue motivating his 1990 paper ``La nouvelle cuisine'' 
as ``the problem
of formulating ... sharply in contemporary physical theory'' ``these
notions, of cause and effect on the one hand, and of correlation on
the other''.  \cite{bell1990}

It is sometimes reported that Bell's local causality
condition is really only a ``no correlation'' requirement, such that
the empirical violation of the resulting inequalities establishes only
``non-local correlations'' (as opposed to non-local causation).  
(See, e.g., Ref.~\onlinecite{gz}.)
But
this is a misconception.  Bell uses the term
``causality'' (e.g., in talking about his ``definition of locally
causal theories'') to highlight that a violation of this condition (by
some theory) means that the theory posits non-local
causal influences, \emph{as opposed to} mere
``non-local correlations.''

It will be clarifying to illustrate this by relaxing a point
that Bell has carefully built into his formulation of local causality,
and showing that violation of the resulting, weakened condition may
still entail \emph{correlations} between space-like separated events,
but no longer implies that there are faster-than-light causal
influences.  Actually, we have done this once already, in the previous section,
when we explained why a violation of Eq.~(\ref{eq-incomplete})
would not (unlike a violation of Eq.~(\ref{locality})) entail any
violation of the causal structure of Figure \ref{fig1}.  We now
consider a second modified version of Bell's criterion.

Consider again the spacetime diagram sketched in Figure 2.  Bell
notes that
\begin{quote}
``It is important that region 3 completely shields off
  from 1 the overlap of the backward light cones of 1 and 2.''
  \cite{bell1990}
\end{quote}
Why is this so important?  
For example, why couldn't we replace region
3 of Figure 2 with a region like that labelled $3^*$ in Figure 3?
This region, just like $3$ in Figure 2, closes off the back light cone
of $1$ and hence -- it might seem -- would be perfectly sufficient for
defining the probabilites associated with $b_1$ in a locally causal 
theory.

\begin{figure}[t]
\begin{center}
\scalebox{.7}{
\scalebox{1} 
{
\begin{pspicture}(0,-1.59)(10.14,1.57)
\definecolor{color82}{rgb}{0.6,0.6,0.6}
\pscircle[linewidth=0.04,dimen=outer](3.3,1.07){0.5}
\usefont{T1}{ptm}{m}{n}
\rput(3.2914062,1.08){$1$}
\psline[linewidth=0.04cm](3.58,1.47)(6.58,-1.53)
\psline[linewidth=0.04cm](3.0,1.45)(0.0,-1.55)
\pscircle[linewidth=0.04,dimen=outer](6.84,1.05){0.5}
\usefont{T1}{ptm}{m}{n}
\rput(6.811406,1.06){$2$}
\psline[linewidth=0.04cm](7.12,1.45)(10.12,-1.55)
\psline[linewidth=0.04cm](6.54,1.43)(3.54,-1.57)
\usefont{T1}{ptm}{m}{n}
\rput(3.3214064,-1.04){$3^*$}
\psbezier[linewidth=0.04,linecolor=color82,tbarsize=0.07055555cm 5.0]{|-}(1.28,-0.47)(1.76,-0.79)(1.96,-0.89)(2.8,-1.01)
\pspolygon[linewidth=0.04,linecolor=White,fillstyle=solid](4.12,-0.81)(4.44,-0.83)(3.9,-1.29)(3.64,-1.33)
\psbezier[linewidth=0.04](1.52,0.01)(2.251087,-0.59)(3.34,-0.93)(5.82,-0.77)
\psbezier[linewidth=0.04](0.72,-0.81)(2.2,-1.55)(4.76,-1.35)(6.24,-1.19)
\psbezier[linewidth=0.04,linecolor=color82,tbarsize=0.07055555cm 5.0]{|-}(5.84,-0.99)(5.16,-1.09)(4.7,-1.11)(3.86,-1.09)
\usefont{T1}{ptm}{m}{n}
\rput(5.0714064,-0.46){$X$}
\end{pspicture} 
}
}
\caption{
Similar to Figure \ref{fig2}, except that region $3^*$ (unlike region
3 of Figure \ref{fig2}) fails to shield off region 1 from the
overlapping backward light cones of regions 1 and 2.  Thus, (following
the language of Figure 2's caption) even full specification of what
happens in $3^*$ \emph{does not make} events in 2 irrelevant for
predictions about 1 in a locally causal theory.
\label{fig3}
}
\end{center}
\end{figure}
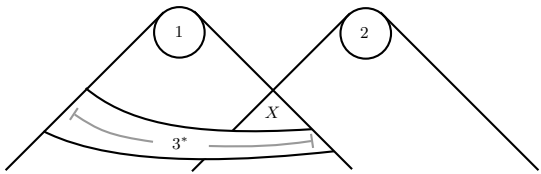

But a more careful analysis shows that a violation of
\begin{equation}
P(b_1 | B_{3^*}, b_2) = P(b_1 | B_{3^*})
\label{localityprime}
\end{equation}
(the same as Eq.~(\ref{locality}) but with region 3 of Figure 2
replaced by
region $3^*$ of Figure 3) does \emph{not} entail any non-local causation.
Here, there is a perfectly \emph{local} causal mechanism by which 
$b_1$ and $b_2$ can be correlated, in a way that isn't ``screened
off'' by conditionalization on $B_{3^*}$, thus violating 
Eq.~(\ref{localityprime}) in a situation which involves no violation of
relativistic local causation.  The mechanism is this:  in a
stochastic theory, an event may occur at the space-time point labelled
``X'' in Figure 3 which was \emph{not determined} by the complete
specification of beables ($B_{3^*}$) in region $3^*$.   But despite not
having been determined by beables in its past, that event really
comes into existence and may in
principle have \emph{effects} throughout its future light cone --
which includes \emph{both} region 1 \emph{and} region 2.  
Event $X$ may, so to speak, broadcast sub-luminal
influences which bring about correlations between $b_1$ and $b_2$,
such that
information about $b_2$ is \emph{not} redundant in regard to defining
what happens in region 1 (even after conditionalizing on $B_{3^*}$).  
Thus we may have a violation of 
Eq.~(\ref{localityprime}) -- i.e., the candidate  theory in question 
could attribute different values to $P(b_1 | B_{3^*}, b_2)$ and
$P(b_1 | B_{3^*})$ --
despite there being, according to the theory in question, no non-local
causation at work.  While Eq.~(\ref{localityprime})
may perhaps be described as some kind of ``no correlations'' condition 
for regions $1$ and $2$, it definitely fails as a ``no causality'' 
condition.  

If we return, however, to the original region 3 (of Figure 2)
which \emph{does} ``completely [shield] off from 1 the overlap of the
backward light cones of 1 and 2'' \cite{bell1990}
it becomes clear that no such
correlation-without-non-local-causality can occur.  Here, if some $X$-like
event (not determined by even a complete specification of beables in
region 3) occurs somewhere in the future light cone of region 3, it
will necessarily fail to lie in the overlapping past light cones of
regions 1 and 2 (which would be necessary for it to in turn locally
influence both of those events).

Bell has
carefully set things up so that a violation of Eq.~(\ref{locality})
entails that there is some non-local causation.  It isn't necessarily
that something in region 2 is causally influencing something in region
1, or vice versa.  It
is always possible that there is some other event, neither in region 1
nor region 2, which was not determined by $B_3$, and which
itself causally influences both $b_1$ and $b_2$.  The point is,
though, that \emph{this} causal influence would have to be non-local
(i.e., would have to violate the special relativistic causal structure
sketched in Figure 1).  \cite{maudlin}

To summarize the point that a violation of Eq.~(\ref{locality})
entails non-local \emph{causation} (rather than mere
\emph{correlations} between space-like separated events) it is helpful
to recall Bell's example of the 
correlation between the ringing of a kitchen alarm
and the readiness of a boiling  egg.  That the alarm rings just as the
egg is finished cooking obviously does not entail or even suggest that
the ringing \emph{caused} the egg to harden.  Correlation does not imply
causality.  As  Bell completes the point,
\begin{quote}
``The
  ringing of the alarm establishes the readiness of the egg.  But if
  it is already given that the egg was nearly boiled a second before,
  then the ringing of the alarm makes the readiness no more certain.''
\cite{bell1990}
\end{quote}
Reading $b_2$ for ``the ringing of the alarm,'' $b_1$ for ``the
readiness of the egg,'' and $B_3$ for ``the egg was nearly boiled a
second before,'' we have a simple intuitive example of
Eq.~(\ref{locality}):  although $b_1$ and $b_2$ may be 
\emph{correlated} such that information about $b_2$ can tell us
something about $b_1$, that information will be redundant (in a locally
causal theory) once $B_3$ is specified.

\subsection{Causality vs. signaling}

One final idea that is often confused with local causality is local
(i.e., exclusively slower-than-light)
\emph{signaling}. \cite{fn:localsignal}   Signaling is,
of course, a certain human activity in which one person transmits
information, across some distance, to another person.  Such
transmission clearly requires a causal connection between the sending
event and the receiving event, but it requires more as well:  namely,
the ability of the people involved to send and receive the
information.  That is, signaling requires some measure of
\emph{control} (over appropriate beables) on the part of the sender,
and some measure of \emph{access} (to appropriate beables) on the part
of the recipient.  

The requirement that theories prohibit the possibility of
faster-than-light signaling -- which incidentally is all that is
imposed in relativistic 
quantum field theory by the requirement that field
operators at spacelike separation commute \cite{bell1976}
-- is thus a much weaker
condition than the prohibition of faster-than-light causal
influences.  Theories can exhibit violations of relativistic 
local causality and yet (because certain beables are inadequately
controllable by and/or inadequately accessible to humans) preclude
faster-than-light signals.  Orthodox quantum mechanics (including
ordinary relativistic quantum field theory) is one example
of such a theory.  Another example is the pilot-wave theory of 
de Broglie and Bohm, in which
\begin{quote}
``...the consequences of events at one place propagate to other places
faster than light.  This happens in a way that we cannot use for
signaling.  Nevertheless it is a gross violation of relativistic
causality.'' \cite{bell1984}
\end{quote}
One of the most prevalent mistakes made by commentators on Bell's
theorem is to conflate local causality with local signaling.  
\cite{fn:signals}
Often
this takes the form of a kind of double-standard in which alternatives
to ordinary QM are dismissed as ``non-local'' (and therefore
unacceptable) on the grounds that they include (either manifestly, 
as in the case of the pilot-wave theory, or just in principle, as 
established by Bell's theorem) ``gross violations of relativistic
causality'' -- but then ordinary QM itself is argued by comparison 
to be perfectly ``local'' (where now only ``local signaling'' is
meant or proved).  Such reasoning, though, is obviously equivocal once one
appreciates that ``local causality'' and ``local signaling'' simply
mean different things.  

Clearly differentiating these two notions does raise the question of
which, after all, SR should be understood to prohibit.  But the idea
that the relativistic causal structure, sketched in Figure 1, should
somehow apply exclusively to this narrowly human activity, seems
highly dubious:
\begin{quote}
``...the `no signaling...' notion rests on concepts which are
desperately vague, or vaguely applicable.  The assertion that `we
cannot signal faster than light' immediately provokes the question:
\begin{center}
Who do we think \emph{we} are?
\end{center}
\emph{We} who can make `measurements', \emph{we} who can manipulate
`external fields', \emph{we} who can `signal' at all, even if not
faster than light?  Do \emph{we} include chemists, or only physicsts,
plants, or only animals, pocket calculators, or only mainframe
computers?'' \cite{bell1990}
\end{quote} 
That is, the idea that SR is compatible with non-local causal
influences (but only prohibits non-local signaling) seems afflicted by
the same problem (reviewed in Section \ref{sec3}) that necessarily
afflicts theories whose formulations involve words like
``observable'', ``microscopic'', ``environment'', etc.  In particular,
the notion of ``signaling'' seems somehow too superficial, too
anthropocentric, to adequately capture the causal structure of 
Figure 1.

\section{Implications of Local Causality}
\label{sec6}

Having reviewed Bell's careful \emph{formulation} of relativistic local
causality, let us now more briefly indicate some of its important
\emph{applications}.

\subsection{Factorization}

In the typical EPR-Bell setup, we have separated observers
(traditionally Alice and Bob) making spin-component measurements
(using, say, Stern-Gerlach devices oriented spatially along the
$\hat{a}$ and $\hat{b}$ directions, respectively) on each of a pair of
spin-entangled particles.  The outcomes of their individual
measurements (manifested in the final location of the particle, or the
position of some pointer, or some fact about some other beable) 
may be denoted by $A$ and $B$ respectively.

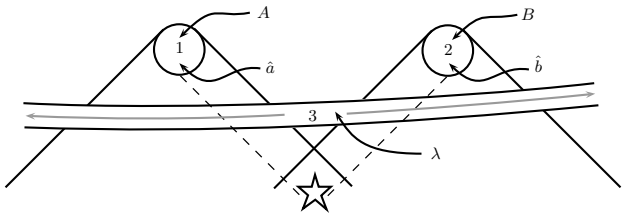
\begin{figure}[t]
\begin{center}
\scalebox{.7}{
\scalebox{1} 
{
\begin{pspicture}(0,-2.0)(11.7,2.0)
\definecolor{color389}{rgb}{0.6,0.6,0.6}
\pscircle[linewidth=0.04,dimen=outer](3.3,1.06){0.5}
\usefont{T1}{ptm}{m}{n}
\rput(4.9014063,1.77){$A$}
\psline[linewidth=0.04cm](3.58,1.46)(6.58,-1.54)
\psline[linewidth=0.04cm](3.0,1.44)(0.0,-1.56)
\pscircle[linewidth=0.04,dimen=outer](8.4,1.04){0.5}
\usefont{T1}{ptm}{m}{n}
\rput(9.921406,1.73){$B$}
\psline[linewidth=0.04cm](8.68,1.44)(11.68,-1.56)
\psline[linewidth=0.04cm](8.1,1.42)(5.1,-1.58)
\pspolygon[linewidth=0.04,linecolor=White,fillstyle=solid](4.8,0.02)(5.2,0.02)(5.58,-0.4)(5.18,-0.44)
\pspolygon[linewidth=0.04,linecolor=White,fillstyle=solid](6.62,0.1)(6.88,0.1)(6.44,-0.36)(6.18,-0.38)
\usefont{T1}{ptm}{m}{n}
\rput(8.171406,-0.91){$\lambda$}
\usefont{T1}{ptm}{m}{n}
\rput(5.0314064,0.73){$\hat{a}$}
\usefont{T1}{ptm}{m}{n}
\rput(10.121407,0.79){$\hat{b}$}
\usefont{T1}{ptm}{m}{n}
\rput(3.3114061,1.09){$1$}
\usefont{T1}{ptm}{m}{n}
\rput(8.4114065,1.05){$2$}
\usefont{T1}{ptm}{m}{n}
\rput(5.831406,-0.19){$3$}
\psbezier[linewidth=0.04,arrowsize=0.05291667cm 2.0,arrowlength=1.4,arrowinset=0.4]{<-}(6.26,-0.12)(6.58,-0.62)(6.8,-0.92)(7.9,-0.92)
\psbezier[linewidth=0.04,arrowsize=0.05291667cm 2.0,arrowlength=1.4,arrowinset=0.4]{<-}(3.3,1.28)(4.0,1.98)(4.0,1.64)(4.64,1.76)
\psbezier[linewidth=0.04,arrowsize=0.05291667cm 2.0,arrowlength=1.4,arrowinset=0.4]{<-}(3.3,0.8)(4.0,0.1)(4.18,0.8)(4.84,0.7)
\psbezier[linewidth=0.04,arrowsize=0.05291667cm 2.0,arrowlength=1.4,arrowinset=0.4]{<-}(8.48,1.24)(9.1,1.84)(9.08,1.6)(9.72,1.72)
\psbezier[linewidth=0.04,arrowsize=0.05291667cm 2.0,arrowlength=1.4,arrowinset=0.4]{<-}(8.42,0.8)(9.1,0.12)(9.28,0.78)(9.94,0.68)
\pspolygon[linewidth=0.04](5.81575,-1.5485486)(5.87775,-1.32)(5.9475,-1.5485486)(6.18,-1.5485486)(5.9965305,-1.716)(6.097755,-1.98)(5.882653,-1.7952)(5.6485715,-1.9668)(5.746,-1.7104372)(5.56,-1.5485486)
\psline[linewidth=0.024cm,linestyle=dashed,dash=0.16cm 0.16cm](3.32,0.56)(5.58,-1.7)
\psline[linewidth=0.024cm,linestyle=dashed,dash=0.16cm 0.16cm](8.38,0.54)(6.12,-1.72)
\pspolygon[linewidth=0.04,linecolor=White,fillstyle=solid](9.68,0.26)(10.02,0.26)(10.36,-0.06)(9.96,-0.1)
\pspolygon[linewidth=0.04,linecolor=White,fillstyle=solid](1.72,-0.04)(1.4,-0.02)(1.02,-0.42)(1.38,-0.42)
\psbezier[linewidth=0.04](0.36,-0.42)(3.94,-0.54)(7.28,-0.38)(11.26,0.0)
\psbezier[linewidth=0.04](0.34,0.04)(3.4,-0.12)(8.4,0.1)(11.18,0.42)
\psbezier[linewidth=0.04,linecolor=color389,arrowsize=0.05291667cm 2.0,arrowlength=1.4,arrowinset=0.4]{<-}(0.38,-0.2)(1.8,-0.26)(3.58,-0.28)(5.3,-0.18)
\psbezier[linewidth=0.04,linecolor=color389,arrowsize=0.05291667cm 2.0,arrowlength=1.4,arrowinset=0.4]{<-}(11.2,0.22)(9.66,0.06)(7.8,-0.08)(6.48,-0.16)
\end{pspicture} 
}
}
\caption{
Space-time diagram illustrating the various beables of  relevance for
the EPR-Bell setup.  (Cf. Bell's diagram in Ref.~\onlinecite{bell1990}.)
Separated observers Alice (in region 1) and Bob
(in region 2) make spin-component measurements (using apparatus settings 
$\hat{a}$ and  $\hat{b}$ respectively) on a pair of spin- or
polarization-entangled particles (represented by the dashed lines).  The
measurements have outcomes $A$ and $B$ respectively.  The state
of the particle pair in region 3 is denoted
$\lambda$.  Note that what we are here calling region 3 extends across
the past light cones of \emph{both} regions 1 and 2.  It thus not only
``completely shields off from 1 the overlap of the backward light
cones of 1 and 2'' \cite{bell1990}, but also vice versa.
Bell's local causality condition therefore requires both that
$\hat{b}$ and $B$ are irrelevant for predictions about the outcome
$A$, and that $\hat{a}$ and $A$ are irrelevant for predictions about
the outcome $B$, once $\lambda$ is specified.
\label{fig4}
}
\end{center}
\end{figure}

The beables pertaining to a given run of the experiment may then be 
cataloged as in Figure 4.  
Roughly, we may think of $\hat{a}$ and
$\hat{b}$ (which live in regions 1 and 2, respectively) 
as referring to the spatial orientations of the two
pieces of measuring apparatus (this being the basis for the notation),
and $\lambda$ (in region 3)
as referring to the state of the particle pair emitted by
the source.  (Note that the phrase ``state of the
particle pair''  should not be taken too seriously; no actual
assumption is made about the existence, for example, of literal
particles.)

Unlike region 3 of Figure \ref{fig2}, region 3 of Figure \ref{fig4}
extends across the past light cone not only of region 1, but of region
2 as well.  It particular, this extended region 3 closes off the past
light cones of \emph{both} regions 1 and 2 and shields \emph{both}
regions from their overlapping past light cones.  A complete
specification of beables in this region 3 will therefore, according to
Bell's concept of local causality, ``make events in  2
irrelevant for predictions about 1'' \cite{bell1990}
\emph{and} will also make events in 1 irrelevant for predictions about 2:
\begin{equation}
P(A|\hat{a},\hat{b},B,\lambda) = P(A|\hat{a},\lambda)
\end{equation}
and
\begin{equation}
P(B|\hat{a},\hat{b},\lambda) = P(B|\hat{b},\lambda).
\end{equation}
From these and the identity
\begin{equation}
P(A,B|\hat{a},\hat{b},\lambda) = P(A|\hat{a},\hat{b},B,\lambda) \cdot
P(B|\hat{a},\hat{b},\lambda) 
\end{equation}
the so-called ``factorization'' of the joint probability for outcomes
$A$ and $B$ immediately follows:
\begin{equation}
P(A,B|\hat{a},\hat{b},\lambda) = P(A|\hat{a},\lambda) \cdot
P(B|\hat{b},\lambda). 
\label{eq-factor}
\end{equation}
This factorization condition is widely recognized in the Bell
literature to be sufficient for the derivation of empirically-testable
Bell-type inequalities.  As Bell notes, however,
\begin{quote}
``Very often such factorizability is taken as the starting point
of the analysis.  Here we have preferred to see it not as the
\emph{formulation} of `local causality', but as a consequence 
thereof.'' \cite{bell1990}
\end{quote}

\subsection{The EPR Argument}

In their famous 1935 paper, Einstein, Podolsky and Rosen argued that a
local explanation for the perfect correlations predicted by quantum
theory (in a certain kind of situation of which the above EPR-Bell
setup is an example) required the existence of locally pre-determined
values for the measurement outcomes. \cite{epr}  Since, as mentioned
in the introduction, ordinary QM contains no such elements of reality,
EPR concluded that ordinary QM (and in particular the wave function) did 
not provide a complete description of physical reality.  They
suggested that an alternative, locally causal theory which \emph{did}
provide a complete description of physical reality might be found.  

Taking for granted that the relevant empirical predictions of quantum
theory are correct, one can summarize the logic of EPR's argument this way:
\begin{equation}
\text{locality} \rightarrow \text{incompleteness}
\label{epr1}
\end{equation}
where ``incompleteness'' means specifically the incompleteness of the
orthodox quantum mechanical description of the particles in question
(in terms of their quantum state alone).  This is of course logically
equivalent to the statement that
\begin{equation}
\text{completeness} \rightarrow \text{non-locality}
\label{epr2}
\end{equation}
which explains why the EPR argument is sometimes characterized as an
argument for the incompleteness of orthodox QM, and sometimes instead
as pointing out the non-locality of that candidate theory.  

In their 1935 paper, EPR appealed to an intuitive notion of local causality
which was not precisely formulated.  It is therefore of interest that
their argument can be recapitulated and made rigorous by using Bell's
formulation of local causality.  It is clarifying to begin with the
EPR argument in the form of Statement \ref{epr2}.  The proof then
consists in simply using the formulated notion of local causality in
its directly-intended way -- namely, to assess whether a
particular candidate theory is or is not local.  

Take again the situation indicated in Figure 4.
Here it is important to appreciate that because of the structure
of region 3 -- and note that it could be extended into a
space-like hypersurface crossing through the region 3 depicted there,
and still satisfy the requirements discussed earlier -- 
the relevant complete specification of beables does not
presuppose that the ``state of the particle pair'' itself must
``factorize'' into two distinct and independent states for the two
particles.  The state can instead be characterized in  a way that is
essentially inseparable, as in ordinary QM, and the argument still
goes through:
``It is notable that in this argument nothing is said about the
locality, or even localizability, of the variable $\lambda$.  These
variables could well include, for example, quantum mechanical state
vectors, which have no particular localization in ordinary
space-time.  It is assumed only that the outputs $A$ and $B$, and the
particular inputs $a$ and $b$, are well localized.''
\cite{bell1981b} 
Let us suppose in particular 
that the ``preparation procedure'' at the particle source
(the star in Figure 4)
gives rise to a pair in (what ordinary QM describes as) 
the spin singlet state
\begin{equation}
|\psi\rangle = \frac{1}{\sqrt{2}}\left( | \uparrow \rangle_1 |\downarrow
 \rangle_2 - |\downarrow \rangle_1 |\uparrow\rangle_2 \right)
\end{equation}
where $|\uparrow\rangle_1$ means that particle 1 is spin-up along
(say) the z-direction, etc.

Suppose also that $\hat{a} = \hat{b} = \hat{z}$,
i.e., both Alice and Bob (freely) choose to measure the spins of the
incoming particles along the z-direction.  Then QM predicts (letting
$A=+1$ denote the result that Alice finds her particle to be spin up,
etc.) that either $A=+1$ and $B=-1$ (with probability 50\%) or that
$A=-1$ and $B=+1$ (with probability 50\%).  

Thus, noting that for orthodox QM $\lambda$ in Figure 4 is simply the
quantum mechanical wave function, we have for example that
\begin{equation}
P(A=+1 | \hat{a}, \lambda) = \frac{1}{2},
\end{equation}
but also that 
\begin{equation}
P(A=+1| \hat{a}, \hat{b}, \lambda, B=-1) = 1,
\end{equation}
in violation of Eq.~(\ref{locality}).  Orthodox QM is not a locally
causal theory:
\begin{quote}
``The theory requires a perfect correlation of [results] on the two
sides.  So specification of the result on one side permits a 100\%
confident prediction of the previously totally uncertain result on the
other side.  Now in ordinary quantum mechanics there just \emph{is}
nothing but the wavefunction for calculating probabilities.  There is
then no question of making the result on one side redundant on the
other by more fully specifying events in some space-time region 3.  We
have a violation of local causality.'' \cite{bell1990}
\end{quote}

As pointed out, Statements \ref{epr1} and \ref{epr2} are logically
equivalent, so it is clear that a locally causal explanation for the
perfect correlations predicted by QM will require a theory with more
(or perhaps just different) beables than just the quantum mechanical
wave function.  But it is possible to show directly from Bell's concept of
local causality that one must in particular posit beables which
pre-determine the outcomes of both measurements.  

To begin with, let us drop the assumption (which applied to ordinary
QM) that it is possible to fully control the state $\lambda$ produced
by the preparation procedure at the source.  Instead, we allow that
$\lambda$ may perhaps take several distinct values from one run of the
experiment to another.  Let us also continue to assume that Alice and
Bob both freely choose to make measurements along the $\hat{z}$ direction.

The argument is then simple:  we have already shown that local
causality entails the factorization of the joint probability for
outcomes $A,B$, once $\lambda$ is specified.
Considering for example the case $A=+1, B=+1$, whose
joint probability vanishes, we therefore have that, for each specific
value of $\lambda$ that might (with nonzero probability) be produced
by the preparation procedure, one of
$P(A=+1 | \hat{a}, \lambda)$ and $P(B=+1|\hat{b},\lambda)$ must vanish.
But since there are only two possible outcomes for each measurement,
each of these possibilities entails that the opposite outcome is 
pre-determined.  For example, 
\begin{equation}
P(A=+1 | \hat{a},\lambda) = 0 \; \rightarrow \; P(A=-1 | \hat{a},
\lambda) = 1
\end{equation}
which means that those particular values of $\lambda$ to which this
applies must ``contain'' or ``encode'' the
outcome $A=-1$ which will then be revealed with certainty if a measurement
along $\hat{a}$ is performed.  It is easy to see that the possible
values of $\lambda$ must therefore fall into two mutually exclusive
and jointly exhaustive categories -- those which encode the
pre-determined outcomes $A=+1$ and $B=-1$, and those which encode the
pre-determined outcomes $A=-1$ and $B=+1$.  \cite{nature}

Indeed, since the measurement axis is assumed to be ``free'' the same
argument will establish that $\lambda$ must encode pre-determined
outcomes for \emph{all} possible measurement directions.  One thus
sees how theories of deterministic hidden variables (or what N. David
Mermin has dubbed ``instruction sets'' \cite{mermin2})
are in fact \emph{required} by local causality.

\subsection{CHSH Inequality}

It is well-known that a Bell-type inequality follows
from the assumption of local deterministic hidden variables or
``instruction sets''.  That theories of this type are actually
required by locality (as explained in the previous sub-section) should
therefore already clarify the seriousness with which Bell took the
idea of a fundamental conflict between SR and the predictions of
QM. This conflict can, however, be brought out in an even more
streamlined way, by deriving 
a Bell-type inequality directly from the factorization of the
joint probability as in Eq.~(\ref{eq-factor}) -- and hence from Bell's
local causality (without any additional discussion of determinism or
pre-determined values).  

Assume that the measurement scenario indicated in Figure \ref{fig4} is
repeated many times, with each setting being
selected ``freely'' or ``randomly'' on each run, from two
possibilities: $\hat{a} \in \{\hat{a}_1,\hat{a}_2\}$, $\hat{b} \in \{
\hat{b}_1,\hat{b}_2\}$.   The procedure which
prepares or creates the ``particles to be measured'' is held fixed for
all runs of the experiment.  As before, this will not necessarily imply that
$\lambda$ is constant for all runs, since the relevant beables may be
less than fully controllable; we will assume, though, that the
distribution of different values of $\lambda$ across the runs can be
characterized by a probability distribution $\rho(\lambda)$.  

Next we define the correlation of outcomes $A$ and $B$ as the
expected value of their product:
\begin{eqnarray}
C(\hat{a},\hat{b}) &=& \int \sum_{A,B} \, A \, B \, P(A|\hat{a}, \lambda) \,
P(B|\hat{b},\lambda) \, \rho(\lambda) \, d\lambda \nonumber \\
& =& \int \bar{A}(\hat{a},\lambda) \, \bar{B}(\hat{a},\lambda) \,
\rho(\lambda) \, d\lambda
\end{eqnarray}
where
\begin{equation}
\bar{A}(\hat{a},\lambda) = P(A=+1|\hat{a},\lambda) -
P(A=-1|\hat{a},\lambda)
\end{equation}
satisfies $|\bar{A}| \le 1$, and similarly for $\bar{B}$.  

Now we consider several combinations of correlations involving
different pairs of settings.  To begin with,
\begin{eqnarray}
&&C(\hat{a}_1,\hat{b}_1) \pm C(\hat{a}_1,\hat{b}_2)  \nonumber \\
&\;& = \int
\bar{A}(\hat{a}_1,\lambda) \, \left( \bar{B}(\hat{b}_1,\lambda) \pm
  \bar{B}(\hat{b}_2,\lambda) \right) \, \rho(\lambda) \, d\lambda
\label{eq-18}
\end{eqnarray}
so that
\begin{eqnarray}
&& \left| \, C(\hat{a}_1,\hat{b}_1) \pm C(\hat{a}_1,\hat{b}_2) \right| 
\nonumber \\
&\;& \le 
\int \left| \bar{B}(\hat{b}_1,\lambda) \pm \bar{B}(\hat{b}_2,\lambda)
\right| \, \rho(\lambda) \, d \lambda .
\label{eq-chsh1}
\end{eqnarray}
Similarly, we have that
\begin{eqnarray}
&& \left| \, C(\hat{a}_2,\hat{b}_1) \mp C(\hat{a}_2,\hat{b}_2) \right|
\nonumber \\
&\;& \le
\int \left| \bar{B}(\hat{b}_1,\lambda) \mp \bar{B}(\hat{b}_2,\lambda)
\right| \, \rho(\lambda) \, d\lambda.
\label{eq-chsh2}
\end{eqnarray}
Adding Equations (\ref{eq-chsh1}) and (\ref{eq-chsh2}), and noting that $|x
\pm y| + |x \mp y|$ is one of $2x$, $-2x$, $2y$, or $-2y$, we have
that 
\begin{eqnarray}
&& \left| \, C(\hat{a}_1,\hat{b}_1) \pm C(\hat{a}_1,\hat{b}_2) \right| + 
\left| \, C(\hat{a}_2,\hat{b}_1) \mp C(\hat{a}_2,\hat{b}_2) \right|
\nonumber \\
&& \; \le 2
\end{eqnarray}
which is the so-called Clauser-Horne-Shimony-Holt (CHSH) inequality. \cite{chsh}
This is 
in essence the relation tested in the experiments, e.g.,
Refs.~\onlinecite{aspect} and \onlinecite{weihs}.  Quantum theory 
predicts that (for appropriate preparations of the two-particle state and 
for appropriate choices of $\hat{a}_1$, $\hat{a}_2$, $\hat{b}_1$, and
$\hat{b}_2$) the left hand side should be $2 \sqrt{2}$ -- more than
40\% larger than the constraint implied by local causality.  And the
experimental results are in excellent agreement with the quantum 
predictions.  

Since the inequality is derived from the local causality condition, 
what follows from the experimental results
is that \emph{any} theory which makes empirically correct
predictions will have to violate the local causality condition.  
As Bell writes, ``The obvious definition of `local causality'
does not work in quantum mechanics, and this cannot be attributed to
the `incompleteness' of that theory.'' \cite{bell1990}

\subsection{The ``Free Variables'' Assumption}

Let us finally return to the assumption that the settings $\hat{a}$
and $\hat{b}$ are ``free'' or ``random''.  Mathematically speaking,
this was the assumption that the probability distribution
$\rho(\lambda)$ for the distribution of possible ``states of the
particle pair'' created by the source is independent of the apparatus
settings $\hat{a}$ and $\hat{b}$.  For example, in deriving
Eq.~(\ref{eq-18}) one assumes that the same probability distribution
$\rho(\lambda)$ characterizes runs in which $\hat{a}_1$ and
$\hat{b}_1$ are measured, as characterizes runs in which $\hat{a}_1$
and $\hat{b}_2$ are measured. 
As Bell writes in support of this assumption,
\begin{quote}
``we may imagine the experiment done on such a scale, with the two
sides of the experiment separated by a distance of order light
minutes, that we can imagine these settings being freely chosen at the
last second by two different experimental physicists....  If these 
last second choices are truly free ..., they are not influenced by the 
variables $\lambda$.  Then the
resultant values for [$\hat{a}$] and [$\hat{b}$] 
do not give any information about
$\lambda$.  So the probability distribution over $\lambda$ does not
depend on [$\hat{a}$] or [$\hat{b}$]...''  \cite{bell1990}
\end{quote}
The real (as opposed to imagined) experiments, however, do not involve
``settings being freely chosen at the last second by two different
experimental physicists'' but instead involve physical random (or
pseudo-random) number generators.  As mentioned earlier, though, this
means that -- at least in principle, for some possible candidate
theories -- a complete description of beables in region 3 of Figure 4
will include not only a complete description of the ``state of the
particle pair'' but also a complete description of whatever physical
degrees of freedom are determining or influencing the eventual
settings $\hat{a}$ and $\hat{b}$ -- making it not only possible but
rather likely that the candidate theory in question should exhibit 
(contrary to the assumption that was made) correlations between what 
we have called $\lambda$ and those settings.  

As suggested earlier, though, one can here appeal to the expectation
that serious candidate theories will posit an enormously large number
of physical degrees of freedom in a spacetime region like 3, only some
tiny fraction of which are actually needed to completely specify the
``state of the particle pair'' -- i.e., the beables which are
physically influenced by the preparation procedure at the source.
There are then innumerable other beables in region 3 which might be
used to determine/influence the apparatus settings.  The expectation
-- more precisely, in the context of the derivation of the locality
inequality, the \emph{assumption} -- is that this is done in some way
such that there are no correlations, no conspiratorial 
pre-established harmonies, between the beables chosen to determine the
apparatus settings and those which encode the ``state of the particle
pair''.  

As Bell acknowledges, therefore, one logical possibility (in the face
of the empirical violations of the CHSH inequality) is that
\begin{quote}
``it is not permissible to regard the experimental settings
[$\hat{a}$] and [$\hat{b}$] in the analyzers as independent 
of the supplementary variables
$\lambda$, in that [$\hat{a}$] and [$\hat{b}$] could be changed 
without changing the
probability distribution $\rho(\lambda)$.  Now even if we have
arranged that [$\hat{a}$] and [$\hat{b}$] are generated by apparently random
radioactive devices, housed in separated boxes and thickly shielded,
or by Swiss national lottery machines, or by elaborate computer
programmes, or by apparently free willed experimental physicists, or
by some combination of all of these, we cannot be \emph{sure} that [$\hat{a}$]
and [$\hat{b}$] are not significantly influenced by the same factors $\lambda$
that influence $A$ and $B$.  But this way of arranging quantum
mechanical correlations would be even more mind boggling than one in
which causal chains go faster than light.  Apparently separate parts
of the world would be deeply and conspiratorially entangled, and our
apparent free will would be entangled with them.''
\cite{bell1981b} 
\end{quote}
And here is another relevant statement from Bell:
\begin{quote}
``An essential element in the reasoning here is that [$\hat{a}$] and
[$\hat{b}$] are free variables.  One can envisage then theories in
which there just \emph{are} no free variables for the polarizer angles
to be coupled to.  In such `superdeterministic' theories the apparent
free will of experimenters, and any other apparent randomness, would
be illusory.  Perhaps such a theory could be both locally causal and
in agreement with quantum mechanical predictions.  However I do not
expect to see a serious theory of this kind.  I would expect a serious
theory to permit `deterministic chaos' or `pseudorandomness', for
complicated subsystems (e.g. computers) which would provide variables
sufficiently free for the purpose at hand.  But I do not have a
theorem about that.'' \cite{bell1990}
\end{quote}
It is sometimes erroneously thought that the ``freedom'' or ``no
conspiracies'' assumption being discussed here does -- or should --
follow from local causality.  For example, in their illuminating
``Exchange [with Bell] on Local Beables'' A. Shimony, M.A. Horne, and
J.F. Clauser criticized Bell's derivation for using (in our notation)
the assumption $\rho(\lambda|\hat{a},\hat{b}) = \rho(\lambda)$ which,
they correctly pointed out, does not follow from local causality.
Bell subsequently clarified that it was a separate assumption, not
supposed to follow from local causality.  As Shimony, Horne, and
Clauser noted, however, the additional assumption seems eminently
reasonable: 
\begin{quote}
``we feel that it is wrong on methodological grounds to worry
seriously about [the possibility of the kind of conspiracy that would
render the assumption inapplicable] if no specific causal linkage
[between the beables $\lambda$ and those which determine the apparatus
settings] is proposed.  In any scientific experiment in which two or
more variables are supposed to be randomly selected, one can always
conjecture that some factor in the overlap of the backward light cones
has controlled the presumably random choices.  But, we maintain,
skepticism of this sort will essentially dismiss all results of
scientific experimentation.  Unless we proceed under the assumption
that hidden conspiracies of this sort do not occur, we have abandoned
in advance the whole enterprise of discovering the laws of nature by
experimentation.''  \cite{exchange} 
\end{quote}
Imagine, for example, an experimental drug trial 
in which patients are randomly selected to receive either the drug or
a placebo.  It is always logically possible that the supposedly random
selections (made, say, by flipping a coin) are in fact correlated with 
some pre-existing facts about the health of patients.  And such
a correlation could skew the results of the trial -- say, resulting in
a statistically significant improvement in the health of the patients
given the genuine drug even though in fact the drug is impotent or
worse.  The suggestion is that, in general, unless there is some plausible
causal mechanism that might conceivably produce the correlations in
question (for example, instead of flipping coins, the drug/placebo
assignments are made on the basis of patients' blood pressures) it is
reasonable to assume that the conspiratorial correlations are absent.

That is, the additional assumption (beyond local causality) which is
needed to derive the CHSH inequality ``is no stronger than one needs
for experimental reasoning generically''. \cite{exchange}  The ``no
conspiracies'' assumption thus falls into the same category as
some other things (for example, the validity of logic, and certain
mathematical operations) which, while used in the derivation, are not
on the table as seriously challengeable.  This
explains why we sometimes do not even bother to mention this
assumption -- as, for example, when writing that the CHSH inequality
follows from Bell's concept of local causality alone.

\section{Summary and Open Questions}
\label{sec7}

We have reviewed J.S. Bell's formulation of relativistic local
causality, including a careful survey of its conceptual background and
a sketch of its most important implications.  We have stressed in
particular that Bell's formulation does not presuppose determinism or
the existence of hidden variables (or any of the other sorts of things
that are sometimes blamed for the empirical violation of Bell-type
inequalities) but instead seems plausibly to just capture the
intuitive idea, widely taken as an implication of special relativity,
indicated in Figure 1 -- namely, the idea that causal influences
cannot propagate faster than light.  And as we have seen, taking now
the ``no conspiracies'' assumption for granted, the
empirically-violated CHSH inequality can be cleanly derived from
Bell's concept of local causality without the need for further
assumptions involving determinism, hidden variables, ``realism'' or
``classicality'' (whatever exactly those ideas mean), etc.  

This hopefully clarifies why Bell disagreed with the widespread
opinion that his theorem -- and the associated experiments -- somehow
vindicate ordinary quantum theory as against hidden variable
alternative theories or somehow vindicate Bohr's philosophy as
against Einstein's.  Instead, the reader can now appreciate
why, for Bell, ``the real problem with quantum theory'' is the
``apparently essential conflict between any sharp formulation and
relativity [, i.e., the] apparent incompatibility, at the deepest level,
between the two fundamental pillars of contemporary theory...''
\cite{bell1984}

It should be noted, however, that the presentation here has been of
\emph{J.S. Bell's} formulation of relativistic local causality.
Although we have argued strongly for its reasonableness, this
particular formulation should not necessarily be regarded as
definitive.

Here we briefly indicate several points on which its
applicability to various sorts of ``exotic'' theories could be
questioned and where a more general or distinct formulation of local
causality might be sought.

For example, one might worry that a theory with a non-Markovian
character (that is, a theory in which causal influences can jump
discontinuously from one time to a later time) could violate Bell's
local causality condition despite positing no strictly
faster-than-light influences.  The idea would be that, in such a
theory, influences could ``hop over'' region 3 of Figure 2, leading to
correlations in regions 1 and 2 but leaving no trace in 3.  
This could seemingly be addressed by requiring that Bell's region 3
extend farther back in time, i.e., cover a region of spacetime so
``thick'' that hopping non-Markovian influences cannot make it across. 
In the limit of arbitrarily large violations of Markovianity, this
would evidently require region 3 to encompass the entire past light
cone of the region 3 pictured in Figure 2.  But this maneuvre
raises further issues:  the more of spacetime that gets included in
region 3, the more one might start to doubt the reasonableness of the
``no conspiracies'' assumption, and the more one might worry that the
condition would fail to detect certain kinds of non-localities (and so
would function merely as a necessary condition for, rather than a
formulation of, locality).  

Similar problems arise when one contemplates the possibility of
theories which posit not only local beables (i.e., those ``being
associated with definite positions in space'' \cite{bell1984b}) 
but also \emph{non-local beables}.  The de Broglie -
Bohm pilot-wave theory is probably the clearest example here:  its
posited ontology includes both particles (which follow definite
trajectories in 3-space and are pre-eminent examples of local beables)
and a guiding wave (which is simply the usual quantum mechanical wave
function but interpreted unambiguously as a beable).  But, for an
$N$-particle system (the universe, say) the wave function is a
function on the $3N$-dimensional configuration space.  It is thus, for
this theory, a non-local beable.  \cite{fn:nomic}

As mentioned previously, Bell's region 3 can be extended into a
space-like hypersurface without spoiling any of the arguments that
have been given in this paper.  We may then generously include, where
Bell's formulation instructs us to use a complete specification of the
\emph{local} beables in region 3, also values for any non-local beables
which, like wave functions, can be associated with hypersurfaces.  And
it is important that, even when including information about non-local
beables in this way, such theories still violate the condition, i.e.,
are diagnosable as non-local.  Still, as formulated, Bell's notion of
local causality seems to presuppose that we are dealing with theories
positing exclusively local beables. \cite{telb}  It can be stretched
to accomodate certain extant theories which posit also non-local
beables, but how to do this with complete generality, and what other
issues (like those encountered above in the case of non-Markovian
theories) may arise in the attempt, remains unclear.  On the other
hand, it is also unclear how seriously one can or should take theories
with non-local beables in the first place. 
In particular:  should such theories even be
considered \emph{candidates} for ``locally causal'' status?  And
perhaps more importantly, could a theory positing non-local beables
possibly count as genuinely consistent with special relativity?  

Such questions will certainly not be answered here.  We raise them
simply to give the reader some sense of the concerns that one might
have about Bell's formulation of local causality.  Their admittedly
exotic character should however help explain why Bell felt driven to
contemplate ``unspeakable'' deviations from conventional wisdom.  In
particular, one can now appreciate how simple everything would become
if we simply dropped the insistence on reconciling the Bell
experiments with ``fundamental relativity'' and instead returned to
the (empirically equivalent) pre-Einstein view according to which
there exists a (hidden) preferred frame of reference.  As explained by
Bell in the quotes given in the Introduction, such a view can easily
accomodate faster-than-light causal influences in a way that
Einsteinian relativity, seemingly, can't.  

Again, though, our goal here is not to lobby for this view, but merely
to explain
Bell's rationale for taking it seriously, as a possibility warranting
careful attention -- not just by philosophers and commentators, but by
physicists interested in addressing the puzzles of yesterday, today,
and tomorrow.

\begin{acknowledgments}
Thanks to Shelly Goldstein, Daniel Tausk, Nino Zanghi, and Roderich
Tumulka for discussions on Bell's formulation of
local causality, and to several anonymous referees for a number of helpful
suggestions on earlier drafts of the paper. 
\end{acknowledgments}


\begin{thebibliography}{00}
\bibitem{fn:localsignal}  What we here call ``local signaling'' is
  also sometimes called ``no signaling'' or ``signal locality''.
\end{thebibliography}
\end{document}